\documentclass{hep99}
\begin{document}

\title{Mirror fermions \\ and the hierarchy problem}

\author{G. Triantaphyllou}

\address{Institute for Theoretical Physics, Technical University Munich \\
James-Franck-Str., D-85748 Garching, Germany
\\[3pt]
E-mail: {\tt georg@ph.tum.de}}

\abstract{The introduction of strongly-interacting mirror fermions with
masses between the
weak scale and 1 TeV could offer a viable alternative to the
Higgs mechanism. The framework provided solves the hierarchy problem 
naturally  and  predicts a rich phenomenology for present and future 
experiments.}
\maketitle

Even though present electroweak precision measurements are more or less
consistent with the standard-model, 
the source of the masses of the known fermions and the weak
vector bosons,  as well as the unitarizing sector of the 
theory, remain unknown. A fundamental scalar field acquiring a non-zero
vacuum expectation value could provide a solution. However, not only 
such a field has not been observed yet, but it is also hard to 
understand why its mass is so much smaller than the Planck scale. 

Efforts to introduce new strongly-interacting fermions as an alternative to
the Higgs mechanism, in the context of technicolor theories for instance,
are difficult to reconcile with precision measurements and unification 
schemes. This recently led \cite{georg1} 
to the introduction of new heavy fermions
with mirror weak-charge assignments, which can overcome these problems
easier. Previous efforts to introduce mirror fermions can be found in 
a review \cite{Maa}; such extensions 
have also been proposed as a solution to
the strong CP problem \cite{Bar}.  
Our interest will be focused here on a particular
version of these fermions which we name ``katoptrons". These  are
differentiated from usual mirror fermions by the fact that they interact
according to  an additional gauge symmetry whose coupling becomes strong
at around 1 TeV, and which gives to all of them dynamical masses 
of that order of magnitude.

In particular, we consider the gauge structure 
$SU(4)_{PS}\times SU(2)_{L} \times SU(2)_{R} \times SU(3)_{2G}$, under which
the standard-model fermions  transform like three copies   
of $({\bf 4,2,1,1})$ and $({\bf {\bar 4},1,2,1})$ and the
katoptrons like  
$({\bf 4,1,2,3})$ and $({\bf {\bar 4},2,1,3})$. At the Pati-Salam 
scale $\Lambda_{PS}$, the gauge symmetry  is reduced according to 
 $SU(4)_{PS} \times SU(2)_{R} \longrightarrow
SU(3)_{C}\times U(1)_{Y}$. As is clearly shown in Fig. 1, this 
easily leads to a unification of all gauge couplings, including the 
$SU(3)_{2G}$ coupling, at a scale $\Lambda_{GUT}$  consistent with
proton life-time bounds \cite{georg2}. Moreover, the $SU(3)_{2G}$
coupling becomes naturally strong at a scale $\Lambda_{M}\approx 1$ TeV.
It allows the formation of katoptron condensates at that scale,
which break the electroweak symmetry dynamically. This provides us
with the first dynamical-symmetry-breaking scenario that post-dicts
correctly the weak scale with natural assumptions (compare with Ref.
\cite{Zee} for instance), and
thus constitutes  a reasonable solution to the hierarchy problem. 

\begin{figure}[h]
\vspace{2.3in}
\includegraphics{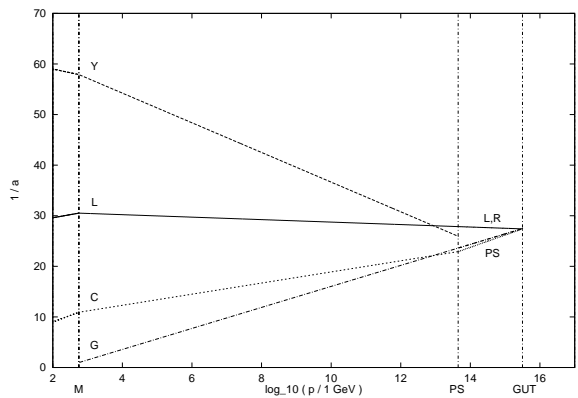}
\caption{The running of the inverse fine-structure constants 
$\alpha^{-1}_{Y,L,C,G}$ and later $\alpha^{-1}_{R,PS}$, corresponding to the
gauge-symmetry breaking channel $SU(4)_{PS} \times SU(2)_{R} \longrightarrow
SU(3)_{C}\times U(1)_{Y}$. The vertical lines, starting from small 
energies, correspond to the scales $\Lambda_{M}=10^{2.75}$ GeV,
$\Lambda_{PS}=10^{13.65}$ GeV and $\Lambda_{GUT}=10^{15.5}$ GeV.}
\end{figure}

The need to generate masses for the ordinary fermions leads us to 
consider scenarios in which the group $SU(3)_{2G}$ breaks after it
becomes strong. Standard-model fermions can then mix with their
mirror partners via gauge-invariant terms in the mass matrix. This can
happen in a way that reproduces correctly not only 
the mass hierarchies but also the mixings of the charged fermions and
neutrinos with each other \cite{georg1},\cite{georg2}. The mechanism
responsible for the breaking of $SU(3)_{2G}$ remains however an important 
open problem.

This breaking is expected to reduce the contribution of the katoptrons
to the $S$ parameter by roughly a factor of 2 \cite{georg3}. This fact, 
together with the existence of light Majorana mirror neutrinos and
negative vertex corrections allow the $S$ parameter to be in agreement
with experimental constraints \cite{georg1},\cite{georg2}. 
Vertex corrections are sufficiently large if the right-handed top-quark
anomalous weak coupling $\delta g^{t}_{R}$ is at
least as large as $\delta g^{b}_{R}$. The latter
is extracted from the deviation of the bottom-quark 
asymmetry $A_{b}$ from its standard-model prediction, 
which being a 2.5 $\sigma$ effect could be the first
experimental indication at hand for the existence of katoptrons
\cite{georg1}. Moreover,
problems with the $\Delta \rho$ parameter can be circumvented since the
top-bottom quark mass hierarchy can be reproduced by gauge-invariant terms
in the mass matrix.

If one regards the $A_{b}$ anomaly as a first indication for the existence
of a heavy mirror sector, one should investigate what signals should be 
expected next in the planned colliders. At the NLC and the Tevatron III, 
the katoptron model forces $V_{tb}$ 
to deviate from its standard-model value,  predicting a 
value around 0.95 \cite{georg1}. The measurement  of 
an anomalous top-quark coupling
$\delta g^{t}_{R}$, which could potentially be even larger than
$\delta g^{b}_{R}$, would also support the katoptron scenario. The LHC could 
produce mirror fermions and their associated scalar resonances
directly, giving however signals  that would hardly be distinguished 
from corresponding fourth-generation or technicolor signals
respectively \cite{georg3}. Note that, since the strong group
$SU(3)_{2G}$ is  eventually broken, it is likely that no vector
resonances are formed and that no $WW$ hard scattering can be observed.

The ultimate proof for the existence of katoptrons would come from a lepton
collider with c.o.m. energies on the order of 4 TeV or higher, like the
muon collider \cite{georg3}. Such a collider would be able to probe
the weak charges of the new fermions directly and thus check their 
mirror-charge assignments. The forward-backward asymmetry of the
katoptrons as a function of collider energy is shown in Fig. 2.

\begin{figure}[ht]
\vspace{2.3in}
\includegraphics{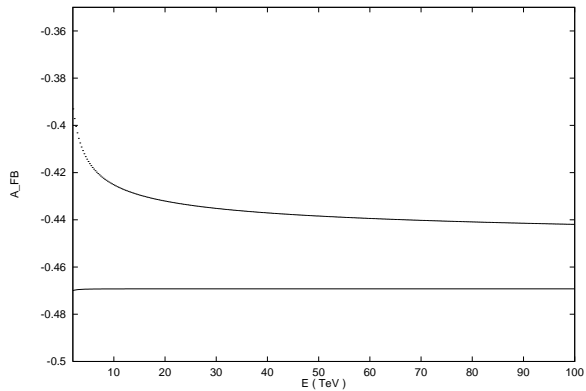}
\caption{The forward-backward asymmetry $A_{FB}$ 
as a function of lepton-collider
c.o.m. energy. The dotted line corresponds to $A_{FB}$ after 
$SU(3)_{2G}$ corrections have been taken into account.}
\end{figure}

It becomes evident from this figure how the
strong $SU(3)_{2G}$ coupling smears out the directional information of 
the out-going katoptrons at low energies, underlining therefore the need
for very powerful lepton colliders. 

To conclude with a unifying perspective, it is useful to remember that some   
representations of gauge groups that arise in superstring theories
include the standard-model fermions not only with their supersymmetric, but
also with their mirror partners. Contrary to what is usually done  
nowadays, the present program is based  on avoiding light scalar
fields by keeping the supersymmetric partners decoupled at 
unification scales and bringing the mirror partners close to the weak scale, 
whose magnitude is, as shown above, an output of the model.

\end{document}